\documentclass[conference]{IEEEtran}
\usepackage{amsmath}
\usepackage{amsfonts}
\usepackage{amssymb}
\usepackage{amsbsy}
\usepackage{graphicx}
\usepackage{times}
\usepackage{rotating}
\usepackage{bm}
\usepackage{default}
\usepackage{color}
\usepackage{cite}
\usepackage{enumerate}

\newtheorem{theorem}{Theorem}
\newtheorem{pro}{Proposition}
\newtheorem{lemma}{Lemma}
\newtheorem{remark}{Remark}

\def\D{D}
\def\L{L}
\def\P{P}
\def\R{R}
\def\Q{Q}
\def\floor{k}
\def\remainder{l}

\def\0v{\boldsymbol{0}}
\def\Iv{\boldsymbol{I}}
\def\av{\boldsymbol{a}}

\def\ev{\boldsymbol{e}}
\def\hv{\boldsymbol{h}}

\def\mv{\boldsymbol{m}}
\def\nv{\boldsymbol{n}}
\def\rv{\boldsymbol{q}}
\def\sv{\boldsymbol{s}}

\def\uv{\boldsymbol{u}}

\def\xv{\boldsymbol{x}}
\def\yv{\boldsymbol{y}}
\def\ybv{{\boldsymbol{\bar y}}}
\def\zv{\boldsymbol{z}}

\def\thv{{\boldsymbol{\theta}}}
\def\Am{\boldsymbol{A}}
\def\Bm{\boldsymbol{B}}
\def\Cm{\boldsymbol{C}}

\def\Jm{\boldsymbol{J}}
\def\Pm{\boldsymbol{P}}

\def\Rm{\boldsymbol{Q}}
\def\Xm{\boldsymbol{X}}

\def\IN{\mathbb{N}}
\def\IC{\mathbb{C}}
\def\IR{\mathbb{R}}
\def\IZ{\mathbb{Z}}
\def\D{\mathcal{D}}
\def\I{\mathcal{I}}

\def\K{\mathcal{K}}
\def\P{\mathcal{P}}

\def\det{\operatorname{det}}
\def\diag{\operatorname{diag}}
\def\rank{\operatorname{rank}}
\def\E{\mathbb{E}}
\begin{document}

\IEEEoverridecommandlockouts

\title{
Noncoherent SIMO Pre-Log via Resolution of Singularities 
}

\author{
\IEEEauthorblockN{Erwin Riegler$^1$, Veniamin I. Morgenshtern$^2$, Giuseppe Durisi$^3$,\\ Shaowei Lin$^4$, Bernd Sturmfels$^4$, Helmut B\"olcskei$^2$}\\
\IEEEauthorblockA{
$^1$Vienna University of Technology, 1040 Vienna, Austria\\
$^2$ETH Zurich, 8092 Zurich, Switzerland\\
$^3$Chalmers University of Technology, 41296 Gothenburg, Sweden\\
$^4$University of California, Berkeley, CA 94720
}
\thanks{The work of Erwin Riegler was partially supported by the WWTF project NOWIRE and carried out while he visited ETH Zurich.}
}

\maketitle
\begin{abstract}
We establish a lower bound on the noncoherent capacity pre-log of a temporally correlated Rayleigh block-fading single-input multiple-output (SIMO) channel. Our result holds for arbitrary rank $\Q$ of the channel correlation matrix, arbitrary block-length $\L>\Q$, and arbitrary number of receive antennas $\R$, and includes the result in Morgenshtern et al. (2010) as a special case. It is well known that the capacity pre-log for this channel in the single-input single-output (SISO) case is given by $1-\Q/\L$, where $\Q/\L$ is the penalty incurred by channel uncertainty. Our result reveals  that this penalty can be reduced to $1/\L$ by adding only one receive antenna, provided that $\L\ge2\Q-1$ and the channel correlation matrix satisfies mild technical conditions.
The main technical tool used to prove our result is Hironaka's celebrated theorem on resolution of singularities in algebraic geometry. 
\end{abstract}

\section{Introduction}
It was shown in \cite{modubo10} that the {\em noncoherent} capacity\footnote{
Noncoherent capacity denotes capacity in the setting where transmitter and receiver know the channel statistics but neither of them is aware of the channel 
realizations.}
pre-log for single-input multiple-output (SIMO) correlated block-fading channels can be larger than the pre-log in the single-input single-output (SISO) case. 
This result was surprising as it disproved a  conjecture in \cite{live04} on the 
 pre-log in the SIMO case being the same as in the SISO case. 
The channel model analyzed in \cite{modubo10} assumes that the fading process is independent across blocks of length $\L$ and temporally correlated within blocks, with the rank of the corresponding $\L\times\L$ channel correlation matrix given by $\Q<\L$ and the 
number of receive antennas $\R=\Q$. 
For this channel model, under additional technical conditions on the channel correlation matrix, a pre-log 
of $(1-1/\L)$ was established in \cite{modubo10}. In contrast, in the SISO case the pre-log is given by $1-\Q/\L$.

The assumption $\R=\Q$ made in \cite{modubo10} is very restrictive, and the proof technique used in \cite{modubo10} 
heavily relies on this assumption. More precisely, the main result in \cite{modubo10} is based on a lower bound on the differential entropy $h(\yv)$ of the channel output signal that is obtained by 
applying a change of variables argument \cite[Lem. 3]{modubo10}. The proof is then completed by showing that 
the expected logarithm of the 
Jacobian  determinant corresponding to this change of variables is finite. For $\R<\Q$, the Jacobian  determinant takes a very involved form making it difficult to say anything about its expected logarithm. The main contribution of this paper is to resolve this problem by introducing a new proof technique based on a result from  algebraic geometry, namely \cite[Th. 2.3]{wa09}, which is a consequence of Hironaka's celebrated theorem on resolution of singularities \cite{hi64,hi64II}. 
Roughly speaking, this result allows us to rewrite any real analytic function 
\cite[Def. 1.6.1]{krpa92} locally as a product of a monomial and a \emph{nonvanishing} real analytic function. The proof of our main result, a lower bound on the pre-log for the correlated block-fading channel with {\em arbitrary} number of receive 
antennas $\R\leq\Q$, is then effected by using this factorization to show 
that the integral of the logarithm of the absolute value of a real analytic function over a compact set is finite, provided that the real analytic 
function is not identically zero. 
This method is very general and could be of independent interest when one tries to show that a certain differential entropy is finite.

We conclude by noting that the main result in this paper shows that 
the pre-log penalty $\Q/\L$ incurred in the SISO case, which is due to channel uncertainty, can be reduced to $1/\L$ by adding only one receive antenna (i.e., by taking $\R=2$), provided that $\L\ge2\Q-1$ and the channel correlation matrix satisfies mild technical conditions. In the limit $\L,\Q\to\infty$ with $\L/\Q$ constant, the $1/\L$ penalty in the SIMO case becomes arbitrarily small, whereas the $\Q/\L$ penalty in the SISO case remains unchanged.

\subsection{Notation}\label{notation}
Finite subsets of the set of natural numbers,  $\I\subset\IN$, are denoted by calligraphic letters and we write $|\mathcal{I}|$ 
for the cardinality of $\mathcal{I}$. We use $[m\!:\!n]$ to designate the set of natural numbers 
$\{m, m+1,\dots,n\}$. 
Uppercase boldface letters denote matrices, lowercase boldface letters designate vectors. 
The superscripts ${}^{\operatorname{T}}$ and ${}^{\operatorname{H}}$ stand for transposition and 
Hermitian transposition, respectively. 
The all-zero matrix of appropriate size is written as $\0v$. 
For a matrix $\Am\in\IC^{M\times N}$, the entry in the $i$th row and $j$th column 
is denoted by $a_{i,j}$ and we write $\av_i^{\operatorname{T}}$ for its $i$th row. If $\I\subseteq [1\!:\!M]$, we denote by 
$\Am_\I\in\IC^{|\I|\times N}$ the submatrix of $\Am$ obtained by retaining all rows $\av_i^{\operatorname{T}}$ of $\Am$ with row index $i\in\I$. 
Similarly, for a vector $\xv\in\IC^{M}$ we denote its $i$th entry by $x_i$. If $\I\subseteq [1\!:\!M]$ we denote 
by $\xv_\I\in\IC^{|\I|}$ the vector obtained by retaining the entries $x_i$ of $\xv$ with $i\in\I$. 
We write $\ev_{i}$ for the $i$th unit vector of appropriate size 
and $\Iv_{M}$ for the identity matrix of size $M\times M$.  
For a vector $\xv$, $\diag(\xv)$ denotes the diagonal matrix that has the entries of $\xv$ in its main diagonal. 
For two matrices $\Am$ and $\Bm$ of arbitrary size,  
$\diag(\Am,\Bm)$ is the $2\times2$ block matrix that has the matrix $\Am$ as upper left block, $\Bm$ as lower right block, and $\0v$ as upper right and lower left block. For $N$ matrices $\Am_1,\dots,\Am_N$, we define 
$\diag(\Am_1,\dots,\Am_N)\triangleq \diag(\diag(\Am_1,\dots,\Am_{N-1}),\Am_N)$. 
We designate the Kronecker product of the matrices $\Am$ and $\Bm$ as 
$\Am\otimes\Bm$; to simplify notation, we use the convention that the ordinary matrix product precedes the Kronecker product, i.e., 
$\Am\Bm\otimes\Cm \triangleq (\Am\Bm)\otimes\Cm$.  
For a function $f$, we write $f\not\equiv 0$ if there exists a vector $\xv$ in the domain of 
$f$ such that $f(\xv)\neq 0$. 
For two functions $f$ and $g$, the notation $f=\mathcal{O}(g)$ means that 
$\lim\sup_{x\to\infty} (|f (x)/g(x)|) < \infty$.
If $x\in\IR$, $\lfloor x\rfloor\triangleq \max\{m\in\IZ\mid m\leq x\}$ and $\lceil x\rceil\triangleq \min\{m\in\IZ\mid m\geq x\}$ with 
$\IZ$ denoting the set of integers. 
The logarithm to the base 2 is written as $\log(\cdot)$. The expectation operator is denoted by $\E[\cdot]$. 
Finally, $\mathcal{CN}(\av,\Cm)$ stands for the distribution of a jointly proper Gaussian random vector with mean $\av$ and covariance matrix $\Cm$.
\section{System model}
We consider a SIMO channel with $\R$ receive antennas. The fading in each component SISO channel follows a correlated block-fading 
model \cite{live04}, with input-output relation for a given block 
\begin{align}\label{eq:model1}
\yv_{m}&=\sqrt{\rho}\diag(\hv_{m})\xv+\nv_{m},\quad m\in [1\!:\!R]
\end{align}
where
$\rho$ denotes the signal-to-noise ratio (SNR), 
$\xv\in\IC^{\L}$ is the transmitted signal vector, 
$\yv_{m}\in\IC^{\L}$ is the received signal vector corresponding to the $m$th receive antenna, and 
$\nv_{m}\sim\mathcal{CN}(\0v,\Iv_{\L})$ is additive noise. 
Finally, $\hv_{m}\sim\mathcal{CN}(\0v,\Rm\Rm^{\operatorname{H}})$ 
is the vector of channel coefficients 
between the transmit antenna and the $m$th receive antenna. Here,\footnote{
When $\Q = \L$, capacity is known to grow double-logarithmically in SNR \cite{lamo03}, and, hence, the pre-log is equal to zero.
If $\R>\Q$ we can always achieve the same pre-log as for $\R=\Q$ by simply using only $\Q$ receive antennas.
} $\Rm\in\IC^{\L\times\Q}$  and $\R\leq\Q\triangleq\rank(\Rm)<\L$.  
 
Without loss of generality, we assume that the row vectors $\rv_i^T$ of $\Rm$ satisfy $\rv_i\neq\0v$ ($i\in[1\!:\!\L]$). 
The vectors $\hv_{m}$ and $\nv_{m}$ are assumed to be mutually independent and independent across 
$m\in [1\!:\!\R]$. It will turn out to be convenient to write the channel-coefficient vector in whitened form as 
$\hv_{m}=\Rm\sv_{m}$, where $\sv_{m}\in\IC^{\Q}$ with $\sv_{m}\sim\mathcal{CN}(\0v,\Iv_{Q})$. Finally, we assume that $\sv_{m}$ and $\nv_{m}$ 
change in an independent fashion from block to block for all $m\in[1\!:\!\R]$. 

Setting 
$\yv^{\operatorname{T}}\triangleq(\yv_{1}^{\operatorname{T}},\dots,\yv_{\R}^{\operatorname{T}})$,  
$\sv^{\operatorname{T}}\triangleq(\sv_{1}^{\operatorname{T}},\dots,\sv_{\R}^{\operatorname{T}})$, 
$\nv^{\operatorname{T}}\triangleq(\nv_{1}^{\operatorname{T}},\dots,\nv_{\R}^{\operatorname{T}})$, and  
$\Xm\triangleq\diag(\xv)$, we can combine the individual input-output relations in 
\eqref{eq:model1}  
into the overall input-output relation
\begin{align}\label{eq:model2}
\yv&=\sqrt{\rho}\ybv+\nv,\quad\text{with}\ \ybv\triangleq(\Iv_{\R}\otimes\Xm\Rm)\sv.
\end{align}
\section{Lower bound on the pre-log}\label{sec:lower}
The capacity of the channel \eqref{eq:model2} is defined as 
\begin{align}\label{eq:capacity}
C(\rho)=(1/\L)\sup_{f(\xv)}I(\xv;\yv)
\end{align}
where $I(\xv;\yv)$ denotes mutual information \cite[p.\,251]{Cover91} and the supremum is taken over all input distributions $f(\xv)$ on $\IC^\L$ that satisfy the average power constraint 
$\E[\|\xv\|^2]\leq\L$. The pre-log is defined as  
$\lim_{\rho\to\infty}(C(\rho)/\log(\rho))$.

The main result of this paper is the following theorem.

\begin{theorem}\label{maintheorem}
Suppose that $\Rm$ satisfies the following 

Property (A): There exists a subset of indices $\K\subseteq[1\!:\!\L]$ with cardinality 
\begin{align}
|\K|\triangleq \min
(\lceil(\Q\R-1)/(\R-1)\rceil,\L)
\end{align}
such that every $\Q$ row vectors of $\Rm_\K$ are linearly independent. 
Then, the capacity of the SIMO channel \eqref{eq:model2} can be lower-bounded as
\begin{align}\label{eq:prelog}
C(\rho) &\geq
\begin{cases}
\ \ \,\,
(1-1/\L)\log(\rho)+\mathcal{O}(1),\ &\text{if}\ \frac{\Q\R-1}{\R-1}\leq\L\\
\R(1-\Q/\L)\log(\rho)+\mathcal{O}(1),\ &\text{else.}
\end{cases}
\end{align}
\end{theorem}
\begin{remark}
The SISO pre-log is $1-\Q/\L$ \cite{live04}.
\end{remark}

\begin{remark}
For $\Q=\R$ Theorem \ref{maintheorem} reduces to \cite[Th. 1]{modubo10}.
\end{remark}
\begin{remark}
Without loss of generality, we will henceforth assume that the set $\K$ in Theorem \ref{maintheorem} is given by $\K=[1\!:\!|\K|]$. This can always 
be achieved by reordering the scalar input-output relations in \eqref{eq:model1}. 
\end{remark}
\begin{remark}
The pre-log is $(1-1/\L)$, provided that $\L\geq2\Q-1$, even if $\R=2$ only. 
\end{remark}
\begin{remark}
Property (A) in Theorem \ref{maintheorem} is not restrictive and is satisfied for a broad class of correlation matrices. 
\end{remark}

\begin{IEEEproof}
Since we are interested in a capacity lower bound, we can evaluate the mutual information in \eqref{eq:capacity} for an appropriate 
input distribution. Specifically, we take the input distribution to have entries $x_i\ (i\in[1\!:\!\L])$ that are independent	and	identically distributed (i.i.d.), zero mean, unit variance, and satisfy $h(x_i) > -\infty$. This implies that \cite[Lem. 6.7]{lamo03}

\begin{align}\label{eq:propx}
\E[\log(|x_i|)]>-\infty,\quad i\in[1\!:\!\L]. 
\end{align}
For example, we can take $x_i \sim \mathcal{CN}(0,1)$. The mutual information $I(\xv;\yv)=h(\yv)-h(\yv\,|\,\xv)$ in \eqref{eq:capacity}, evaluated for any input distribution satisfying these constraints, 
 is then lower-bounded as follows. We first upper-bound $h(\yv\,|\,\xv)$ according to \cite[Eq. (8)]{modubo10}
\begin{align}\label{eq:boundhyx}
h(\yv\,|\,\xv)
&\leq\Q\R\log(\rho)+\mathcal{O}(1)
\end{align}
and then lower-bound $h(\yv)$ as in \cite[Eq. (12)]{modubo10}
\begin{align}
h(\yv)\label{eq:boundhy}
&\geq (\Q\R+\L-\alpha)\log(\rho)+h(\Pm\ybv\,|\,\xv_\P) + c
\end{align}
where $c$ is a constant that is independent of $\rho$, $h(\Pm\ybv\,|\,\xv_\P)$ is independent of $\rho$, 
$\P\triangleq [1\!:\!\alpha]$ with $\alpha\in [1\!:\!\L]$, and 
\begin{align}
\Pm&\triangleq \diag({(\Iv_{\L})}_{\I_1}  ,\dots,{(\Iv_{\L})}_{\I_{\R}})\label{eq:Pm}
\end{align}
for sets $\I_1,\dots,\I_{\R}\subseteq [1\!:\!\L]$ satisfying 
\begin{align}\label{eq:dimension}
\sum_{i\in[1:\R]}|\I_{i}|
=\Q\R+\L-\alpha.
\end{align}
The set $\P$ can be interpreted as a set of pilot positions \cite{modubo10}.  
Combining \eqref{eq:boundhyx} and \eqref{eq:boundhy}, the capacity lower bound 
in \eqref{eq:prelog} is established by choosing 
\begin{align*}
\alpha =
\begin{cases}
1,&\text{if}\ (\Q\R-1)/(\R-1)\leq\L\\
\Q\R-(\R-1)\L,&\text{else},
\end{cases}
\end{align*}
provided that we can find sets $\I_1,\dots,\I_{\R}$ such that 
$h(\Pm\ybv\,|\,\xv_\P)>-\infty$. The remainder of the paper is devoted to identifying such a choice for $\I_1,\dots,\I_{\R}$ and 
proving that the corresponding differential entropy $h(\Pm\ybv\,|\,\xv_\P)$ is, indeed, finite. 
The main idea is to choose the sets $\I_1,\dots,\I_{\R}$ such that  
$h(\Pm\ybv\,|\,\xv_\P)$ can be related to $h(\sv,\xv_{\D}) = h(\sv)+h(\xv_{\D})$   
with $\D\triangleq [\alpha+1\!:\!\L]$ through a deterministic 
one-to-one mapping. 
The quantity  
$h(\sv)+h(\xv_\D)$ is much easier to deal with than $h(\Pm\ybv\,|\,\xv_\P)$. 

Condition \eqref{eq:dimension}  
implies that the 
mapping 
\begin{equation}\label{eq:mapping}
(\sv,\xv_{\D})\mapsto \Pm\ybv=\Pm(\Iv_{\R}\otimes\Xm\Rm)\sv
\end{equation}
is between two vector spaces of the {\em same} dimension $\Q\R+\L-\alpha$, which is a necessary condition for this mapping to be   
one-to-one. Note that the RHS of \eqref{eq:mapping} also depends on $\xv_\P$, which is, however, taken to be fixed, 
reflecting the fact that the pilot symbols are known to both transmitter and receiver. 
Any dependence on $\xv$ will henceforth implicitly mean a dependence on $\xv_\D$ only. 
We set $\I_\R\triangleq [1\!:\!\L]$ and shall choose $\I_1,\dots,\I_{\R-1}\subseteq [1\!:\!\L]$ as follows: 
\begin{enumerate}[(a)]
\item If $(\Q\R-1)/(\R-1)>\L$, we set $\I_{1}\triangleq\dots\triangleq\I_{\R-1}\triangleq [1\!:\!\L]$. 
\item If $(\Q\R-1)/(\R-1)\leq\L$, we let  
\begin{equation}\label{eq:sets}
\begin{split}
\I_m&\triangleq [1\!:\!\Q+\floor+1],\quad m\in [1\!:\!\R-\remainder-1]\\
\I_m&\triangleq [1\!:\!\Q+\floor+2],\quad m\in [\R-\remainder\!:\!\R-1]
\end{split}
\end{equation}
with  
$\floor\triangleq \lfloor(\Q-\R)/(\R-1)\rfloor$ and $\remainder\triangleq \Q-\R -\floor(\R-1)$. 
\end{enumerate}
Now let 
\begin{align}\label{eq:Jacobian1}
\Jm(\sv,\xv_{\D})&\triangleq \frac{\partial \Pm\ybv}{\partial (\sv,\xv_{\D})} 
\end{align}
be the Jacobian 
of the mapping in \eqref{eq:mapping}. 
If this mapping is one-to-one on $\IC^{\Q\R+\L-\alpha}$ almost everywhere (a.e.), 
we  
can apply the change-of-variables theorem for integrals \cite[Th. 7.26]{ru87} in combination with \cite[Th. 7.2]{FrGr10}
and find that 
\begin{align*}
h(\Pm\ybv\mid\xv_\P) 
&=
h(\sv, \xv_{\D})+ 2\,\E[\log(|\det(\Jm(\sv,\xv_{\D}))|)].
\end{align*}
The proof is then concluded  by establishing that the mapping in \eqref{eq:mapping} is one-to-one a.e. and   
\begin{align}\label{eq:toshow}
\E[\log(|\det(\Jm(\sv,\xv_{\D}))|)]
>-\infty.
\end{align}
This requires an in-depth analysis of the Jacobian in \eqref{eq:Jacobian1}, which will be carried out in the next section. 
\end{IEEEproof}

\section{Properties of the Jacobian}\label{Jacobian}
The following lemma provides important insights into the structure of the determinant of the Jacobian in \eqref{eq:Jacobian1}. 

\begin{lemma}\label{lammaJac}
The Jacobian in \eqref{eq:Jacobian1} can be decomposed as
\begin{align}\label{eq:factorization}
\Jm(\sv,\xv_{\D})
&=\Jm_1(\xv_{\D})\Jm_2(\sv)\Jm_3(\xv_{\D})
\end{align}
where 
\begin{align}
\Jm_1(\xv_{\D})\label{eq:J1}
&\triangleq \Pm(\Iv_{\R}\otimes\Xm)\Pm^{\operatorname{T}}\\
\Jm_2(\sv)
&\triangleq \Pm[\Iv_{\R}\otimes\Rm\mid \av_{\alpha+1}\mid \dots\mid \av_{\L}]\label{eq:J2}\\
\Jm_3(\xv_{\D})
&\triangleq \diag(\Iv_{\Q\R},\diag(\xv_\D)^{-1})\label{eq:J3}
\end{align}
with 
\begin{align}\label{eq:ai}
\av_i\triangleq (\Iv_{\R}\otimes\diag(\ev_{i})\Rm)\sv,\quad i\in [1\!:\!\L].
\end{align}
\end{lemma}

\begin{IEEEproof} 
The lemma follows by noting that   
\begin{align*}
\frac{\partial\ybv}{\partial x_i}
&= \frac{\partial}{\partial x_i}\Big(\sum_{j\in [1:\L]}x_j(\Iv_{\R}\otimes\diag(\ev_{j})\Rm)\sv \Big)\\
&=\av_i,\quad i\in [1\!:\!\L]\\
\frac{\partial\ybv}{\partial \sv}
&=\Iv_{\R}\otimes\Xm\Rm.
\end{align*}
\end{IEEEproof}
Based on \eqref{eq:propx}, we can conclude that\footnote{We assume that $x_j\neq 0$ for all $j\in\P$.} 
\begin{align*}
&\E[\log(|\det(\Jm_1(\xv_{\D}))\det(\Jm_3(\xv_{\D}))|)]\nonumber\\
&=
\sum_{j\in\P}
\R\log(|\xv_j|)
+\!\!\!\!\sum_{i\in [1:\R-1]}\sum_{j\in\I_{i}\setminus\P}
\E[\log(|\xv_j|)] >-\infty.
\end{align*}
To conclude the proof of \eqref{eq:toshow} it therefore remains to show that $\E[\log(|\det(\Jm_2(\sv))|)]>-\infty$. 
Direct computation reveals that each vector $\Pm\av_i$ in \eqref{eq:J2} with $i\in\I_{\R}\setminus\I_{\R-1}$ contains only one nonzero element, 
which is given by $\rv^{\operatorname{T}}_{i}\sv_\R$. 
Applying the Laplace formula \cite[p.7]{hojo85} to 
$\Jm_2(\sv)$ in \eqref{eq:J2} therefore yields the decomposition  
\begin{align}\label{eq:splitJ2}
|\det(\Jm_2(\sv))|
&=|\det(\Jm_4(\sv))|\prod_{i\in\I_\R\setminus\I_{\R-1}}|\rv^{\operatorname{T}}_{i}\sv_\R|
\end{align}
with 
\begin{align}\label{eq:J4}
\Jm_4(\sv)&\triangleq \Pm_1[(\Iv_{\R}\otimes\Rm)\mid \av_{\alpha+1}\mid \dots\mid \av_{|\I_{\R-1}|}]
\end{align}
and 
\begin{align*}
\Pm_1&\triangleq \diag((\Iv_{\L})_{\I_1},\dots, (\Iv_{\L})_{\I_{\R-1}}, (\Iv_{\L})_{\I_{\R-1}}).
\end{align*}
The expectation of the logarithm of the second term on the RHS in \eqref{eq:splitJ2} is finite because $\sv_{\R}\sim\mathcal{CN}(\0v,\Iv_{Q})$. 
It remains to show that $\E[\log(|\det(\Jm_4(\sv))|)]>-\infty$.  
This is the most technical part in the proof of Theorem \ref{maintheorem} and can be accomplished by applying methods from algebraic geometry, namely Theorem \ref{TheoLin} in 
Appendix \ref{resolution}, which is a consequence of Hironaka's celebrated theorem on resolution of singularities  
\cite{hi64,hi64II}. 
A direct proof would require showing that the expected log of the determinant of the (high-dimensional) matrix $\Jm_4(\sv)$ is finite, which seems exceedingly difficult. Hironaka's theorem drastically simplifies the proof as it tells us 
that $\det(\Jm_4(\cdot))\not\equiv 0$ implies that 
$\E[\log(|\det(\Jm_4(\sv))|)]>-\infty$. We start by noting that 
$\det(\Jm_4(\sv))$ is a homogeneous polynomial in $s_1,\dots,s_{\Q\R}$ of degree 
 $D\triangleq |\I_{\R-1}|-\alpha$, i.e., 
\begin{align}\label{eq:homogeneous2}
\det(\Jm_4(\lambda\sv))=\lambda^{D} \det(\Jm_4(\sv)),\quad\forall\lambda\in\IC, 
\end{align}  
which allows us to  apply the following proposition (for  
 $M=\Q\R$ and $D$ defined above):

\begin{pro}\label{proHomogeneous}
Let $g$ be a homogeneous polynomial in $s_1,\dots,s_M$ of degree $D\in\IN_0$ with 
$\sv\sim\mathcal{CN}(\0v,\Iv_M)$. Then $g\not\equiv 0$ implies that  
\begin{equation}\label{eq:1}
\E[\log(|g(\sv)|)]> -\infty.
\end{equation}
\end{pro}
\begin{IEEEproof}
Writing $\sv$ as $\sv=\|\sv\|(\sv/\|\sv\|)$ and using the fact that  $g$ is a homogeneous polynomial of degree $D$,  
we can upper-bound the absolute value of the expectation in \eqref{eq:1} by 
\begin{align*}
|\E[\log(|g(\sv)|)]|\leq 
D\underbrace{\E[|\!\log(\|\sv\|)|]}_{\triangleq A}+
\underbrace{\E[|\!\log(|g(\sv/\|\sv\|)|)|]}_{\triangleq B}.
\end{align*}
Then \cite[Lem. 6.7]{lamo03} together with $\sv\sim\mathcal{CN}(\0v,\Iv_M)$ implies that $A<\infty$. 
Introducing polar coordinates \cite[p. 55]{mu82} $r\in \IR_+$
and 
\[
\thv\triangleq (\theta_1,\dots,\theta_{2M-1})^{\operatorname{T}}\in \Delta\triangleq [0,\pi]^{2M-2}\times[0,2\pi] 
\]
 for the complex vector $\sv\in\IC^M$,  
we can 
further upper-bound $B$ according to 
\begin{align*}
B&\leq \int_0^\infty \exp(-r^2) r^{2M-1} \operatorname{d}\!r 
\times
\int_{\Delta}
|\log (|f(\thv)|)| \operatorname{d}\!\thv
\end{align*}
where $f$ is obtained from $g$ by changing to polar coordinates $r\in\IR_+, \thv\in\Delta$.
Note that $f$ is a real analytic function \cite[Def. 1.6.1]{krpa92}, $g\not\equiv 0$ implies that $f\not\equiv 0$, 
and we are integrating $|\log(|f(\cdot)|)|$ over a compact set $\Delta$. We can therefore apply Theorem \ref{TheoLin} in Appendix 
\ref{resolution} to conclude that $B<\infty$. 
\end{IEEEproof}

It remains to show that $\det(\Jm_4(\cdot))\not\equiv 0$ for our specific choice of sets $\I_1,\dots,\I_{\R-1}$, which will be proved in the following lemma: 

\begin{lemma}\label{lemmazero}
Property (A) in Theorem \ref{maintheorem} implies that $\det(\Jm_4(\cdot))\not\equiv 0$.
\end{lemma}
\begin{IEEEproof}
See Appendix \ref{lemmazeroproof}
\end{IEEEproof}

In summary, we proved that \eqref{eq:toshow} holds provided that Property (A) in Theorem \ref{maintheorem} is satisfied. 
It turns out that $\det(\Jm_4(\cdot))\not\equiv 0$ also implies that 
the mapping in \eqref{eq:mapping} is one-to-one a.e. on $\IC^{\Q\R+\L-\alpha}$.  The proof, which is along the lines of the proof of \cite[Lem. 2]{modubo10}, is omitted due to space limitations. 
This completes the proof of Theorem \ref{maintheorem}.

\appendix
\subsection{Resolution of singularities}\label{resolution}

In this appendix, we show how Hironaka's theorem on resolution of singularities can be used to prove that $\int_{\Delta\subset\IR^M}|\log(|f(\xv)|)|d\xv<\infty$ provided that $f\not\equiv 0$ is a real analytic function and $\Delta$ is a compact 
set. 

We start by defining notation that will be used in this appendix. 
Let $C^M(\xv,\epsilon)\triangleq \{\yv\in\IR^M\mid |x_i-y_i|<\epsilon,\ \forall\ i\in [1\!:\!M]\}$ 
denote the open cube with side length 
$2\epsilon$ centered at $\xv$. For $\xv\in\IC^M$ and $\mv\in\IN_0^M$ let 
$\xv^{\mv}\triangleq x_1^{m_1}\dots x_M^{m_M}$. 
If $\Sigma$ is a subset of the image of a map $f$ then 
$f^{-1}(\Sigma)$ denotes the inverse image of $\Sigma$.  

The following lemma is an immediate consequence of a modified version of Hironaka's theorem \cite[Th. 2.3]{wa09}. This modified version originally 
appeared in \cite{at70}. The main point of this lemma is that it allows us to rewrite any real analytic function \cite[Def. 1.6.1]{krpa92} $f\not\equiv 0$ locally as a product of a monomial and a nonvanishing real analytic function. 

\begin{lemma}\label{Lemmasing}
Let $f$ be a real analytic function from a neighborhood $\Omega\subseteq \IR^D$ of $\xv\in\IR^D$ to $\IR$. Suppose that $f(\xv)=0$.
Then, there exists a triple $(\Sigma, \Xi, \phi)$, where
\begin{enumerate}[(a)]
\item $\Sigma\subseteq \Omega$ is an open set in $\IR^D$ with $\xv\in \Sigma$,
\item $\Xi$ is a $D$-dimensional real analytic manifold \cite[Def. 2.10]{wa09}, 
\item $\phi: \Xi\to \Sigma$ is a real analytic map
\end{enumerate}
that satisfies the following conditions:
\begin{enumerate}[(1)]
\item The map $\phi$ is proper, i.e., the inverse image of any compact set is compact.
\item The map $\phi$ is a real analytic isomorphism between $\Xi\setminus (f\circ \phi)^{-1}(0)$ and $\Sigma\setminus f^{-1}(0)$.
\item For each point $P\in \Xi$, there exists a coordinate chart $\{\Xi_{P},\varphi_{P}\}$ such that 
$P\in \Xi_{P}\subseteq \Xi$, $\varphi_{P}\!:\! C^D(\0v,\epsilon_{P})\to \Xi_{P}$ is a real analytic isomorphism for some $\epsilon_{P} >0$ with 
$\varphi_{P}(\0v)=P$, 
\[
|f(\phi\circ\varphi_P(\uv))| 
= h_{P}(\uv)\uv^{\mv_{P}},
\quad \forall\uv\in C^D(\0v,\epsilon_{P})
\]
where $h_{P}$ is a nonvanishing real analytic function on $C^D(\0v,\epsilon_{P})$ and $\mv_{P}\in\IN_0^D$, and the determinant of the Jacobian of the mapping 
$\phi\circ\varphi_P$ satisfies 
\[
\det\Big(\frac{\partial (\phi\circ\varphi_P(\uv))}{\partial \uv}\Big)
= g_{P}(\uv)\uv^{\nv_{P}},\quad\forall\uv\in C^D(\0v,\epsilon_{P})
\]
where $g_{P}$ is a nonvanishing real analytic function on $C^D(\0v,\epsilon_{P})$ and $\nv_{P}\in\IN_0^D$.
\end{enumerate}
\end{lemma}
\begin{IEEEproof} 
The main idea is to apply \cite[Th. 2.3]{wa09} to the function 
$\tilde f(\zv)\triangleq f(\zv+\xv),\ \forall\zv\in\Omega-\xv$. 
We omit the details due to space limitations.
\end{IEEEproof}

We are now in a position to state the theorem that is needed to prove $B<\infty$ in the proof of 
Proposition \ref{proHomogeneous}.  

\begin{theorem}\label{TheoLin}
Let $f\not\equiv 0$ be a real analytic function on an open set $\Omega\subset\IR^D$.
Then 
\begin{equation}\label{eq:Lineq}
\int_{\Delta} |\log(|f(\xv)|)| \operatorname{d}\!\xv < \infty
\end{equation}
for all compact sets $\Delta\subset \Omega$.
\end{theorem}
\begin{IEEEproof}
Let $\xv\in\Delta$. If $f(\xv)=0$
then 
Lemma \ref{Lemmasing} 
implies that there exists a triple
 $(\Sigma_{\xv},\Xi_{\xv},\phi_{\xv})$ where 
$\Sigma_{\xv}\subseteq\Omega$ is an open set containing $\xv$, $\Xi_{\xv}$ is a real analytic manifold, and 
$\phi_{\xv}\!:\!\Xi_{\xv}\to \Sigma_{\xv}$ is a proper real analytic map. Moreover, 
for each $P\in \Xi_{\xv}$ there exists a  coordinate chart $\{\Xi_{\xv,P},\varphi_{\xv,P}\}$ such that 
$\Xi_{\xv,P}=\varphi_{\xv,P}(C^D(\0v,\epsilon_{\xv,P}))$, $\varphi_{\xv,P}(\0v)=P$, and 
\begin{equation}
\begin{split}\label{eq:chvar}
|f\circ\phi_{\xv}\circ\varphi_{\xv,P}(\uv)|=h_{\xv,P}(\uv)\uv^{\mv_{\xv,P}}\\
\det\Big(\frac{\partial \phi_{\xv}\circ\varphi_{\xv,P}(\uv)}{\partial \uv}\Big) = g_{\xv,P}(\uv)\uv^{\nv_{\xv,P}}
\end{split}
\end{equation}
for all $\uv\in  C^D(\0v,\epsilon_{\xv,P})$, where $g_{\xv,P}$ and $h_{\xv,P}$ are nonvanishing real analytic functions 
on $C^D(\0v,\epsilon_{\xv,P})$. We can choose $\epsilon_{\xv,P}$ sufficiently small so that $g_{\xv,P}$ and $h_{\xv,P}$ 
are bounded on $C^D(\0v,\epsilon_{\xv,P})$. 
If $f(\xv)\neq 0$ the existence of a triple  $(\Sigma_{\xv},\Xi_{\xv},\phi_{\xv})$ with the properties specified above is guaranteed by 
taking $\Xi_{\xv}=\Sigma_{\xv}$ sufficiently small such that $f$ does not vanish on $\Sigma_{\xv}$ and by setting $\phi_{\xv}$ to be the identity map.

Now for each $\xv \in \Delta$, we choose an open neighborhood
$\Sigma'_{\xv}$ and a compact neighborhood $\Delta_{\xv}$ such
that $\xv \in \Sigma'_{\xv} \subset \Delta_{\xv} \subset
\Sigma_{\xv}$. Since $\Delta$ is a compact set, there exists a finite
set of vectors $\{\xv_{1},\dots,\xv_{N}\}$ in $\Delta$ such that
\begin{align*}
\Delta\,\,\subset \bigcup_{i\in [1:N]} \Sigma'_{\xv_i}\subset
\bigcup_{i\in [1:N]} \Delta_{\xv_i}.
\end{align*}
For each $i\in[1\!:\!N]$, set  $\Delta_{i}\triangleq
\Delta_{\xv_{i}}$,  $\Sigma_{i}\triangleq \Sigma_{\xv_{i}}$,
$\Xi_{i}\triangleq \Xi_{\xv_{i}}$, and
$\phi_{i}\triangleq\phi_{\xv_{i}}$. Since 
the mapping $\phi_{i}\!:\!\Xi_{i}\to \Sigma_{i}$ is proper,  
each set 
${\phi_{i}}^{-1}(\Delta_{i})\subset \Xi_{i}$ is a compact set. Therefore, there exists a finite set of points 
$\{P_{1},\dots,P_{M_{i}}\}$ in $\Xi_{i}$ such that 
\begin{align}\label{eq:compactinU}
{\phi_{i}}^{-1}(\Delta_{i})\subset\bigcup_{j\in[1:M_{i}]}\Xi_{i,j} 
\end{align}
with $\Xi_{i,j}\triangleq \Xi_{\xv_{i},P_{j}}$. Since \eqref{eq:compactinU} holds for all $i\in[1\!:\!N]$, 
we can upper-bound the integral in \eqref{eq:Lineq} as follows:
\begin{align*}
&\int_{\Delta} |\log (|f(\xv)|)|\operatorname{d}\!\xv
\leq \sum_{i\in [1:N]}\int_{\Delta_{i}} |\log (|f(\xv)|)|\operatorname{d}\!\xv\\ 
&\leq 
c_1\!\sum_{i\in [1:N]}
\sum_{j\in [1:M_{i}]}
\int_{C^D(\0v,\epsilon_{i,j})}
|\uv^{\nv_{i,j}}
\log\big(|\uv^{\mv_{i,j}}|\big)|\operatorname{d}\!\uv
+c_2\\&
<\infty
\end{align*}
where $c_1,c_2>0$ are positive real numbers, $g_{i,j}, h_{i,j}$ are bounded nonvanishing real analytic functions on 
$C^D(\0v,\epsilon_{i,j})$, $\mv_{i,j}, \nv_{i,j}$ are vectors of nonnegative integers, and we changed variables according to 
\eqref{eq:chvar}.
\end{IEEEproof}

\subsection{Proof of Lemma \ref{lemmazero}}\label{lemmazeroproof}
We present a proof for $\alpha=1, \remainder\neq 0$ and skip the (simpler) cases $\alpha > 1$ and $\alpha=1, \remainder=0$.

Suppose that $\alpha=1$ and $\remainder\neq 0$. 
We can write $\Jm_4(\sv)$ in \eqref{eq:J2} as 
$\Jm_4(\sv)=[\diag(\Rm_{\I_1},\dots, \Rm_{\I_{\R-1}}, \Rm_{\I_{\R-1}})\ \Am]$ 
with 
$\Am^{\operatorname{T}}\triangleq [\Am_1^{\operatorname{T}}\ \dots\ \Am^{\operatorname{T}}_{\R}]$ 
and $\Am_m$ defined as 
\begin{align*}
\Am_m&\triangleq 
\begin{pmatrix}
0&0&0&0\\
\rv^{\operatorname{T}}_{2}\sv_m&0&0&0\\
\vdots&\ddots&0&0\\
0&\dots&\rv^{\operatorname{T}}_{\Q+\floor+1}\sv_m&0
\end{pmatrix},
\  m\in [1\!:\!\R-\remainder-1]\\
\\
\Am_m&\triangleq 
\begin{pmatrix}
0&0&0&\\
\rv^{\operatorname{T}}_{2}\sv_m&0&0\\
\vdots&\ddots&0\\
0&\dots&\rv^{\operatorname{T}}_{\Q+\floor+2}\sv_m
\end{pmatrix},\ \ m\in [\R-\remainder\!:\!\R].
\end{align*}
Property (A) in Theorem \ref{maintheorem} implies that for  
arbitrary subsets 
\begin{align*}
\K_m&\subseteq [2\!:\!\Q+k+1],\quad m\in [1\!:\!\R-\remainder-1]\\
\K_m&\subseteq [2\!:\!\Q+k+2],\quad m\in [\R-\remainder\!:\!\R]
\end{align*}
with $|\K_m|=\Q-1$,  we can find vectors $\sv_m\in\IC^{\Q}$ ($m\in[1\!:\!\R]$) such that 
\begin{enumerate}[(a)]
\item $\rv^{\operatorname{T}}_{j}\sv_m=0$ for all vectors $\rv^{\operatorname{T}}_{j}$ with $j\in\K_m$;
\item $\rv^{\operatorname{T}}_{j}\sv_m\neq 0$ for all vectors with $j\in \K\setminus\K_m$. 
\end{enumerate}
This implies that for each choice of such sets $\K_m$ ($m\in[1\!:\!\R]$), there exists a set of vectors  $\sv_m\in\IC^{\Q}$ ($m\in[1\!:\!\R]$) such that 
the number of nonzero elements $n_m$ in each matrix $\Am_m$ satisfies 
\begin{equation}\label{eq:ni}
\begin{split}
n_m&=\floor+1,\quad m\in [1\!:\! \R-\remainder-1]\\
n_m&=\floor+2,\quad m\in [\R-\remainder\!:\!\R].
\end{split}
\end{equation}
Moreover, we have 
\begin{align*}
\sum_{m\in[1:\R]}n_m
&=\Q+\floor+1
\end{align*}
which implies that we can choose the subsets $\K_m$ and the vectors $\sv_m$ ($m\in[1\!:\!\R]$) such that each column of 
$\Am$ contains precisely one nonzero element. 
Applying the Laplace formula \cite[p. 7]{hojo85} iteratively, we therefore get 
\begin{align*}
|\det(\Jm_4(\sv))|
&=c
\prod_{m\in [1:\R]}
|\det(\Rm_{\K_m\cup\{1\}})|>0
\end{align*}
where $c$ is a positive constant and we used Property (A) in Theorem \ref{maintheorem} in the last step. 
\endproof
\renewcommand{\baselinestretch}{1.05}\small\normalsize

\bibliography{references}
\bibliographystyle{IEEEtran}
\end{document}